\documentclass[amssymb,amsmath,prd,twocolumn,superscriptaddress,nofootinbib]{revtex4-1}

\usepackage{graphicx}
\usepackage{bm}
\usepackage{amssymb,amsmath}
\usepackage{mathrsfs}
\usepackage{latexsym}
\usepackage{color}
\usepackage[normalem]{ulem} 
\usepackage{dcolumn}
\usepackage[colorlinks=true,citecolor=blue,urlcolor=blue]{hyperref}
\usepackage[usenames,dvipsnames]{xcolor}

\allowdisplaybreaks

\newcommand{\CITA}{\affiliation{Canadian Institute for Theoretical Astrophysics, 60 St. George Street, Toronto, Ontario, M5S 3H8, Canada}}
\newcommand{\CCA}{\affiliation{Center for Computational Astrophysics, Flatiron Institute, 162 5th Ave, New York, NY 10010, USA}}
\newcommand{\LIGOlabMIT}{\affiliation{LIGO Laboratory, Massachusetts Institute of Technology, 185 Albany St, Cambridge, MA 02139, USA}}
\newcommand{\MKI}{\affiliation{MIT-Kavli Institute for Astrophysics and Space Research, 77 Massachusetts Ave, Cambridge, MA 02139, USA}}
\newcommand{\Austin}{\affiliation{Theory Group, Department of Physics, University of Texas at Austin, Austin, TX 78712, USA}}

\definecolor{azgreen}{rgb}{0.03,0.47,0.19}
\definecolor{chmagenta}{rgb}{0.54, 0.17, 0.88}

\newcommand{\nn}{\nonumber}

\newcommand{\BF}{{\mathcal B}}

\begin{document}

\title{On combining information from multiple gravitational wave sources}

\author{Aaron Zimmerman}
\CITA
\Austin
\author{Carl-Johan Haster}
\CITA
\LIGOlabMIT
\MKI
\author{Katerina Chatziioannou}
\CITA
\CCA

\date{\today}

\begin{abstract}
In the coming years, advanced gravitational wave detectors will observe signals from a large number of compact binary coalescences.
The majority of these signals will be relatively weak, making the precision measurement of subtle effects, such as deviations from general relativity, challenging in the individual events. 
However, many weak observations can be combined into precise inferences, if information from the individual signals is combined in an appropriate way.
In this study we revisit common methods for combining multiple gravitational wave observations to test general relativity, namely
(i) multiplying the individual likelihoods of beyond-general-relativity parameters and (ii) multiplying the Bayes Factor in favor of general relativity from each event.
We discuss both methods and show that they make stringent assumptions about the modified theory of gravity they test. 
In particular, the former assumes that all events share the same beyond-general-relativity parameter, while the latter
assumes that the theory of gravity has a new unrelated parameter for each detection.
We show that each method can fail to detect deviations from general relativity when the modified theory being tested violates these assumptions.
We argue that these two methods are the extreme limits of a more generic framework of hierarchical inference on hyperparameters that characterize the underlying distribution of single-event parameters.
We illustrate our conclusions first using a simple model of Gaussian likelihoods, and also by applying parameter estimation techniques to a simulated dataset of gravitational waveforms in a model where the graviton is massive.
We argue that combining information from multiple sources requires explicit assumptions that make the results inherently model-dependent.
\end{abstract}

\maketitle
  
\section{Introduction}

Two years after the first direct detection of gravitational waves (GWs) from the coalescence of two black holes (BHs), GW astronomy has transitioned into a booming field with eleven confirmed detections to date~\cite{LIGOScientific:2018mvr}. 
The rate of detections is only expected to increase in the coming years as the GW detector network becomes more sensitive and grows in number~\cite{Aasi:2013wya}. 
This has lead to increasing interest in the problem of combining information from multiple GW sources to strengthen inferences that might be inconclusive from single events. 
Examples include tests of general relativity (GR)~\cite{TheLIGOScientific:2016pea}, the measurement of the Hubble constant~\cite{Chen:2017rfc,Feeney:2018mkj}, as well as inference about population properties~\cite{LIGOScientific:2018mvr,LIGOScientific:2018jsj}.

Within the framework of Bayesian Inference, combining information from multiple sources is usually formulated in two complementary ways. 
The first involves computing the posterior probability distribution of parameters that characterize the entire population, such as the coupling constant of a beyond-GR theory or the shape of the mass distribution of stellar BHs. 
The second amounts to computing the \emph{Bayes Factor} (BF), the ratio of the evidences of two competing models that describe the population.

In the context of testing GR with GWs, two approaches are commonly used. 
The first involves computing the BF in favor of GR compared to beyond-GR theories for each individual signal and then multiplying the individual BFs~\cite{Li:2011cg,Li:2011vx,Agathos:2013upa,Meidam:2014jpa}. 
The second approach involves computing the posterior probability distribution of some parameter that quantifies the deviation from GR and then multiplying the individual likelihoods together in order to draw stronger inferences on this parameter~\cite{DelPozzo:2011pg,Meidam:2014jpa,Ghosh:2016qgn,TheLIGOScientific:2016pea,Ghosh:2017gfp,Brito:2018rfr}. 
This parameter might be a fundamental constant of the theory that is common between all binary systems, such as the Compton wavelength of the graviton, or a parameter that is expected to depend on the specific properties of each binary system, such as the modifications to the post-Newtonian expansion introduced within the parametrized post-Einsteinian (ppE) framework~\cite{Yunes:2009ke,Chatziioannou:2012rf}.

In this study, we revisit both approaches in order to discuss the underlying assumptions about the properties of the beyond-GR theory they test. 
Unsurprisingly, we find that multiplying the individual likelihoods of a beyond-GR parameter is appropriate only for theories where this parameter is identical for every member of the population. 
Meanwhile, multiplying the BFs is appropriate only when each the beyond-GR parameter is different for every new system observed.
Through a simple model of Gaussian likelihoods we show that both approaches lead to faulty conclusions if applied to a theory of gravity that does not follow their respective assumptions.
In particular, we may fail to detect deviations from GR when using incorrect assumptions about the underlying modification to GR.

The two assumptions above, namely that parameters are \emph{equal} or \emph{completely uncorrelated} are arguably extreme assumptions that generic theories of gravity are not required to obey.
Instead it is more reasonable to expect the middle ground of parameters that are related to each other through an underlying population distribution of the beyond-GR parameters.
This scenario is akin to, for example, the expectation that the masses of stellar-mass BHs in binaries are drawn from an underlying population mass distribution.
We show that hierarchical analyses where one parametrizes and infers the underlying distribution can be interpreted as an intermediate case that encompasses the two extremes of multiplying likelihoods or BFs.

Our results raise the issue of the model-dependency of combining information from multiple signals, something particularly critical for tests of GR with GWs. 
We argue that even though model-independent tests are feasible for single sources, this notion of model-independent tests does not carry over to multiple detections. 
The rest of the paper presents the details of our argument. 
Section~\ref{sec:analytical} studies the assumptions inherently made when combining inferences and illustrates these ideas using a simple model of Gaussian likelihoods. 
Section~\ref{sec:graviton} presents a concrete example in terms of measuring the Compton wavelength of the graviton. 
Section~\ref{sec:conclusions} summarizes our main conclusions.

\section{Assumptions when combining inferences}
\label{sec:analytical}

In this section we outline assumptions made when using two common methods for combining inferences from multiple detections in order to test deviations from relativity.
Each of these is an extreme case of a more general framework, where the parameters controlling the deviation of a set of signals from the predictions of GR are drawn from some common (and generally unknown) distribution.
In order to better illustrate these ideas and how constraints of modified theories scale with the number of signals analyzed, we use a simple model of Gaussian likelihoods.
We then illustrate how the two common methods of combining inferences can fail to detect deviations from relativity when synthetic signals violate the underlying assumptions of each method.

\subsection{Model comparison}
\label{Sec:Definitions}

Consider two competing hypotheses $H_1$ and $H_2$ that we wish to test with data $D$. The BF in favor of $H_1$ is given by
\begin{align}
\BF & = \frac{p(D | H_1)}{p(D|H_2)} \,,
\end{align}
where $p(D|H_i)=\int p(\bm{\theta}_i) p(D|\bm{\theta}_i, H_i) d \bm{\theta}_i$ is the evidence for $H_i$, $\bm{\theta}_i$ are the model parameters, $p(\bm{\theta}_i)$ is the prior on the parameters, and $p(D|\bm{\theta}_i, H_i)$ is the likelihood.

We assume that $H_1$ and $H_2$ are nested models, i.e.~$\bm{\theta}_2= \{ \bm{\theta}_1,\bm{\lambda}\}$ and $p(D|\bm{\theta}_1, H_1)=p(D|\bm{\theta}_1,\bm{\lambda}=0,H_2)$. 
The posterior for the extra parameters $\bm{\lambda}$ is
\begin{align}
p(\bm{\lambda}|D,H_2)&=\frac{p(\bm{\lambda}| H_2)p(D|\bm{\lambda}, H_2)}{p(D|H_2)} \nn
\\
=&\frac{p(D| \bm{\lambda},H_2)}{\int p(\bm{\lambda}',H_2) p(D| \bm{\lambda}',H_2) d \bm{\lambda}'} \,.
\end{align}
and using the Savage-Dickey density ratio~\cite{dickey1971} the BF simplifies to 
\begin{align}
\BF=\frac{p(\bm{\lambda}=0|D,H_2)}{p(\bm{\lambda}=0| H_2)}\,.\label{SDdef}
\end{align}
From here on we suppress the explicit dependence on $H_1$ and $H_2$.

As an example, assume that $H_2$ has only one additional parameter, $\lambda$, for which the likelihood is Gaussian with a median $\mu$ and standard deviation $\sigma$,
\begin{align}
p(D|\lambda) & = c e^{-(\lambda - \mu)^2/(2\sigma^2)} \,,
\end{align}
and that the prior is uniform over the width $\Delta \lambda$, $p(\lambda) = 1/\Delta \lambda$.
The BF is then
\begin{align}
\BF & = \frac{1}{\sqrt{2 \pi } }e^{-\mu^2/(2 \sigma^2)} \frac{\Delta \lambda}{\sigma}\,.
\end{align}
The last term, which is greater than $1$, is the usual Occam's penalty for the fact that $H_2$ has additional parameters and is penalized as compared to $H_1$.

Consider now two independent measurements, for example two GW events with $D=\{D_1,D_2\}$, and that the independent events have a parameters $\lambda_1$ and $\lambda_2$, respectively. 
The posterior for the deviation parameters after both measurements is
\begin{align}
\label{eq:TwoEventPosterior}
p(\lambda_1,\lambda_2|D) & = \frac{p(\lambda_1,\lambda_2) p(D| \lambda_1,\lambda_2)}{\int p(\lambda_1',\lambda_2') p(D|\lambda_1',\lambda_2') d \lambda_1' d \lambda_2'} \nn
\\
&= \frac{p(\lambda_1,\lambda_2) p(D_1| \lambda_1)p(D_2| \lambda_2)}{\int p(\lambda_1',\lambda_2') p(D_1|\lambda_1')p(D_2|\lambda_2') d \lambda_1' d \lambda_2'} \nn
\\
&= \frac{p(\lambda_1,\lambda_2) p(D_1| \lambda_1 )p(D_2| \lambda_2)}{\int p(\lambda_1')p(\lambda_2'|\lambda_1') p(D_1|\lambda_1')p(D_2|\lambda_2') d \lambda_1' d \lambda_2'} ,
\end{align}
where to obtain the second line we have factorized the likelihood by assuming that each measurement is independent, and to obtain the third line we have used the identity $p(\lambda_1,\lambda_2)=p(\lambda_1)p(\lambda_2|\lambda_1)$.

Crucially, before we can proceed, we need to specify $p(\lambda_2|\lambda_1)$, which quantifies the relation between $\lambda_1$ and $\lambda_2$. 
Notice that $\lambda_1$ and $\lambda_2$ are parameters describing different systems, but measurement of one can influence our prior belief on the value of the other. 
This is similar to, for example, the fact that measurement of the mass of one binary affects our prior expectation for the mass of a future binary as the masses are linked through a common astrophysical mass distribution~\cite{LIGOScientific:2018jsj}.
We consider two extreme cases which are commonly used in the literature: one where each GW signal shares a common parameter $\lambda$, and one where the two additional parameters $\lambda_i$ are unrelated.
We also discuss the most general case, where the properties of each individual signal are related by some common, underlying distribution.

\subsection{Case 1: Common parameter and multiplying likelihoods}
\label{case1}

In the case where we have identified a common parameter for all GWs then $\lambda_1=\lambda_2\equiv\lambda$ and
\begin{align}
p(\lambda_2|\lambda_1) = \delta(\lambda_1-\lambda_2).
\end{align}
Now in order to make sense of the posterior, we marginalize over $\lambda_2$, then simplify the result as
\begin{align}
\label{eq:CommonPosterior}
p(\lambda|D)&=p(\lambda_1|D)=\int p(\lambda_1,\lambda_2|D)d \lambda_2 \nn
\\
&=\int \frac{p(\lambda_1,\lambda_2) p(D_1| \lambda_1)p(D_2| \lambda_2)}{\int p(\lambda_1')p(\lambda_2'|\lambda_1') p(D_1|\lambda_1')p(D_2|\lambda_2') d \lambda_1' d \lambda_2'}d \lambda_2\nn
\\
& = \frac{p(\lambda)p(D_1| \lambda)p(D_2| \lambda)}{\int p(\lambda') p(D_1|\lambda')p(D_2|\lambda') d \lambda' } .
\end{align}
This equation shows that this case is equivalent to multiplying the individual likelihoods for $\lambda$ obtained by each event, then multiplying by one instance of the common prior and normalizing the resulting probability distribution. 
In other words, this is the standard situation of measuring a common parameter using multiple, independent trials, where the posterior distribution from the previous measurement can be used as a prior for the next measurement.

To clearly illustrate the result we specialize to uniform priors and Gaussian likelihoods, so that
\begin{align}
p(D_i|\lambda) & = c_i e^{-(\lambda - \mu_i)^2/(2\sigma_i^2)} \,.
\end{align}
The product of the two Gaussian likelihoods is proportional to another Gaussian with 
\begin{align}
\sigma_f^2 & = \frac{\sigma_1^2 \sigma_2^2}{\sigma_1^2 + \sigma_2^2} \,, & \mu_f & = \frac{\mu_1 \sigma_2^2 + \mu_2 \sigma_1^2}{\sigma_1^2 + \sigma_2^2} \,.
\end{align}
With this the BF is
\begin{align}
\label{eq:CoherentTwoEvents}
\BF & = e^{-\mu_f^2/(2 \sigma_f^2)} \frac{\Delta \lambda}{\sqrt{2 \pi } \sigma_f}\,,
\end{align}
but now $\sigma_f$ is smaller than the individual standard deviations, and $\mu_f$ is a weighted average of the maximum likelihood value for $\lambda$ coming from each measurement.
The Occam's penalty factor is modified from the case of a single measurement only by the change in $\sigma$. 

From here is it straightforward to generalize to $N$ events. 
For each of the $N$ events, we measure the deviation parameter $\lambda_i$.
Since the events are independent, the overall likelihood again factorizes, and only the prior $p(\lambda_1, \dots, \lambda_N)$ links the inferences.
Repeating the proceeding steps for multiple common parameters, where all $\lambda_i = \lambda$, we see that the posterior is proportional to the product of the individual likelihoods and a single instance of the prior,
\begin{align}
p(\lambda|D) \propto p(\lambda) \prod_i p(D_i,\lambda) \,.
\end{align}

To show how this influences the BF, we again assume the likelihoods are proportional to Gaussians, and note that for the multiplication of $N$ Gaussians, the result is again proportional to a Gaussian with parameters
\begin{align}
\label{eq:Nvar}
\sigma^2 & = \left(\sum_i^N \frac{1}{\sigma_i^2} \right)^{-1} \,, & \mu & = \sigma^2 \sum_i^N \frac{\mu_i}{\sigma_i^2} \,,
\\
\label{eq:Nnorm}
\ln C & = -\frac 12 \sum_i^N \frac{\mu_i^2}{\sigma_i^2} + \frac{\mu^2}{2\sigma^2}  \,.
\end{align}
Now the BF is given by Eq.~\eqref{eq:CoherentTwoEvents} with the above values of $\sigma$ and $\mu$ (the factors of $C$ cancel in the numerator and denominator).
To get a clearer picture of how this scales with $N$, we assume all events have the same $\mu_i$ and $\sigma_i$, in which case
\begin{align}
\BF & = \sqrt{N} e^{-N \mu_i^2/(2 \sigma_i^2)} \frac{\Delta \lambda}{\sqrt{2 \pi } \sigma_i} \,.\label{BFcommon}
\end{align}
In the case where $H_1$ is correct and so $\mu_i =0$, our BF grows with $\sqrt{N}$. 
In reality even in the case where GR is correct and $\mu_i =0$, the specific noise realization of the detector noise will cause variations in $\mu_i$ with a standard deviation of $\sigma_i$. 
Below we numerically show that even if we include this variation in our results the $\sqrt{N}$ scaling is preserved on average. 

\subsection{Case 2: Unrelated parameters and multiplying BFs}
\label{case2}

In the opposite extreme, we assume that all systems have their own additional parameter, so that we learn nothing
about $\lambda_2$ from measuring $\lambda_1$.
Returning briefly to the case where we have only two events, in Eq.~\eqref{eq:TwoEventPosterior} we set
\begin{align}
p(\lambda_2|\lambda_1) = p(\lambda_2),
\end{align}
and then Eq.~\eqref{eq:TwoEventPosterior} becomes
\begin{align}
p(\lambda_1,\lambda_2|D) & = \frac{p(\lambda_1,\lambda_2)p(D_1| \lambda_1)p(D_2| \lambda_2)}{\int p(\lambda_1')p(\lambda_2'|\lambda_1') p(D_1|\lambda_1')p(D_2|\lambda_2') d \lambda_1' d \lambda_2'}\nn
\\
& = \frac{p(\lambda_1) p(\lambda_2) p(D_1| \lambda_1)p(D_2| \lambda_2)}{\int p(\lambda_1') p(\lambda_2') p(D_1|\lambda_1')p(D_2|\lambda_2') d \lambda_1' d \lambda_2'} \nn
\\
& = \frac{p(\lambda_1) p(D_1| \lambda_1)}{\int p(\lambda_1') p(D_1|\lambda_1') d \lambda_1' }\frac{p(\lambda_2)p(D_2| \lambda_2)}{\int p(\lambda_2') p(D_2|\lambda_2') d \lambda_2'} .
\end{align}
From this posterior and using Eq.~\eqref{SDdef} we see that the total BF in this case reduces to the product of the individual BFs.
Generalizing to $N$ measurements and under the assumption that all $\lambda_i$ are unrelated, the total BF is given by the product of the individual BFs, a simplification sometimes used in the literature~\cite{Li:2011cg,Li:2011vx,Agathos:2013upa,Meidam:2014jpa}.

If we specialize once again to Gaussian likelihoods, we find 
\begin{align}
\label{eq:BFTwoInd}
\BF & = C e^{-\mu_f^2/(2 \sigma_f^2)} \frac{\Delta \lambda_1}{\sqrt{2 \pi } \sigma_1}\frac{ \Delta \lambda_2}{\sqrt{2 \pi } \sigma_2}\,,
\end{align}
where $C$ is the constant prefactor of the multiplied Gaussians, see Eq.~\eqref{eq:Nnorm}.
We see when comparing to Eq.~\eqref{eq:CoherentTwoEvents} that $H_2$ is penalized by two Occam's factors, one for each $\lambda_i$, rather than the single factor for the extra parameter $\lambda$. 
This makes sense, since now $H_2$ includes two new independent parameters.

Meanwhile, for $N$ measurements where the $\lambda_i$ are all unrelated to each other we have that the product of BFs is
\begin{align}
\BF & = C e^{- \mu^2/(2 \sigma^2)} \prod_i \frac{\Delta \lambda_i}{\sqrt{2 \pi \sigma_i}}\,.\label{BFrandom}
\end{align}
We see that we have $N$ Occam's factors which penalize $H_2$.
Again to get a concrete picture we let all the likelihoods be equal, in which case $C = 1$, and all prior are ranges equal, so that
\begin{align}
\BF & =  e^{-N \mu_i^2/(2 \sigma_i^2)}  \left( \frac{\Delta \lambda}{\sqrt{2\pi} \sigma_i} \right)^N \,.
\end{align} 
In the case where $H_1$ is the true theory, $\mu_i = 0$ and this BF grows exponentially with $N$.

We conclude that this method leads in general to larger BFs in favor of GR and tighter constraints compared to multiplying the likelihoods. However, this method assumes that the theory of gravity has $N$ coupling constants with respect to GR, one for every signal detected. 

\subsection{General case: Deviation parameters drawn from a common distribution}
\label{sec:GeneralModel}

So far we have described two extreme cases for computing joint inference for nested hypotheses.
In each case, certain assumptions were used to simplify the conditional probability on the deviation parameters $p(\lambda_1, \dots, \lambda_N)$.
The most general case would allow for a nontrivial distribution for $p(\lambda_1, \dots, \lambda_N)$, though not all distributions 
are physically realistic for GW signals.
The various datasets are still assumed to be independent of each other, but now the parameters $\lambda_i$ are drawn from a common distribution, similar for example to the expectation that BBH masses are drawn from a common mass distribution.
We can then assume that the prior distributions for $\lambda_1$ and $\lambda_2$ can be related by a common set of additional {\it hyperparameters}~\cite{2004AIPC..735..195L}. 

As a concrete example for an underlying distribution, we return to our Gaussian theme and consider $\lambda_i$ drawn from a Gaussian distribution centered on $M$, with standard deviation $\alpha$,
\begin{align}
p(\lambda_i|M,\alpha) & = \frac{1}{\sqrt{2\pi}\alpha}e^{-(\lambda_i - M)^2/(2\alpha^2)}\,.
\end{align}
With this, the posteriors of the individual events will be modified Gaussians with widths $\tilde \sigma_i$ and means $\tilde \mu_i$,
\begin{align}
p(\lambda_i) p(D_i|\lambda_i) & = \frac{1}{2 \pi \sigma_i \alpha_i} \tilde C e^{-(\lambda_i - \tilde \mu_i)^2/(2 \tilde \sigma_i^2)} \,,
\end{align}
where
\begin{align}
\tilde \sigma_i^2 &= \frac{\sigma_i^2 \alpha^2}{\sigma_i^2 + \alpha^2} \,, & \tilde \mu_i & = \frac{\mu_i \alpha^2 + M \sigma_i^2}{\sigma_i^2 + \alpha^2} \,,
\end{align}
and the overall normalization factor is
\begin{align}
\ln \tilde C_i & = \frac 12 \left(\frac{\tilde \mu_i^2}{\tilde \sigma_i^2} - \frac{\mu_i^2}{\sigma_i^2} - \frac{M^2}{\alpha^2}  \right)   = - \frac{(\mu_i - M)^2}{2(\sigma_i^2 + \alpha^2)}\,.
\end{align}
With these definitions, we find that the evidences of the individual posteriors are
\begin{align}
\int p(\lambda_i) p(D_i|\lambda_i) d \lambda_i = \frac{1}{\sigma_i} \frac{\tilde \sigma_i \tilde C_i}{\sqrt{2 \pi} \alpha} \,.
\end{align}
The combined BF can be shown to be similar to Eq.~\eqref{eq:BFTwoInd}, with the numerator unchanged and the denominator modified as above.
The result is
\begin{align}
\BF & = \frac{C}{\tilde C_1 \tilde C_2} e^{-\mu_f^2/(2 \sigma_f^2)} \frac{\alpha}{\tilde \sigma_1} \frac{\alpha}{\tilde \sigma_2} \,.
\end{align}
Note that as we hold $\alpha$ fixed and vary $\sigma_i$, we always have $\alpha \geq \tilde \sigma_i$.

The above analysis gives the BF in the case where $H_2$ is associated with a particular choice of $M$ and $\alpha$.
If we are interested in the BF for a hypothesis where the $\lambda_i$ are drawn from some Gaussian with unknown $M$ and $\alpha$, then we should marginalize over these hyperparameters, using some prior $p(M, \alpha)$.
This leads to
\begin{align}
\BF & = Ce^{-\mu_f^2/(2 \sigma_f^2)} 
\left( \int  \frac{\tilde C_1 \tilde \sigma_1}{\alpha} \frac{ \tilde C_2 \tilde \sigma_2}{\alpha} 
p(M, \alpha) dM d\alpha \right)^{-1} \,.
\end{align}

For example, our first case where all the $\lambda_i$ are equal corresponds to drawing the $\lambda_i$ from a Gaussian with zero width $\alpha=0$, centered on some unknown $\lambda = M$. 
To accomplish this we set $p(M,\alpha) = p(M) \delta(\alpha)$.
Assuming a flat hyperprior on $M$, $p(M) = 1/(\Delta M)$, we integrate in order to recover the BF of Eq.~\eqref{eq:CoherentTwoEvents}, with the substitution $\Delta \lambda \to \Delta M$.

The second case, where the $\lambda_i$ are treated as unrelated parameters, is given by taking $\alpha$ large and enforcing a cutoff on the prior range.
In this case, we fix $M$ to some value within the prior range and let $\alpha \to \infty$, which we accomplish by setting $p(M, \alpha)$ to the appropriate delta functions and integrating over them.
Then the BF reduces to Eq.~\eqref{eq:BFTwoInd}.

The framework above, where each $\lambda_i$ are drawn from some common distribution controlled by a set of hyperparameters, is the most general framework for systems which do not interact.
We have shown how each of the two common methods for combining multiple GW events arise out of simple limits from this framework.

\subsection{Combining synthetic observations from multiple events}
\label{sec:CombSynth}
 
The expressions for the BF presented in Eqs.~\eqref{BFcommon} and~\eqref{BFrandom} were obtained under specific assumptions about the parameters $\lambda_i$ that describe the deviation from GR. 
In order to study the implications of indiscriminately applying these formulas to situations and datasets that violate these assumptions, we return to our toy model of Gaussian likelihoods and numerically compute the BFs by either multiplying the likelihoods, Eq.~\eqref{BFcommon}, or by multiplying the BFs of the individual measurements together, Eq.~\eqref{BFrandom}.
To simulate a mock population of measurements for $\lambda_i$ we draw the mean $\mu_i$ of the Gaussian likelihood from various distributions and for simplicity set the standard deviation to $\sigma_i=1$. 
We also assume a flat prior on $\lambda_i$.
 
We focus on $5$ example cases for the distribution of $\mu_i$: 
\begin{enumerate}
\item $\mu_i=0+ {\cal{N}}(0,\sigma_i=1)$: This corresponds to the case that GR is correct, but the posteriors have 
a scatter around the true value of $\mu_i=0$ due to the detector noise realization. Here ${\cal N}(\mu, \sigma)$ indicates a normal distribution of mean $\mu$ and standard deviation $\sigma$.
\item $\mu_i=1+ {\cal{N}}(0,\sigma_i=1)$: In this case GR is not correct, but the true theory of gravity predicts
the same beyond-GR parameter for each event, $\mu_i=1$, plus a scatter due to detector noise. This corresponds
to the situation discussed in Sec.~\ref{case1}.
\item $\mu_i\in [-1,1]+ {\cal{N}}(0,\sigma_i=1)$: Here GR is not correct and the true theory of gravity predicts
beyond-GR parameters that are unrelated to each other, plus a scatter due to detector noise. This theory breaks the assumptions of Sec.~\ref{case1}.
\item  $\mu_i \in [-4,4] + {\cal{N}}(0,\sigma_i=1)$: This is the same as 3 but here we draw $\mu_i$ from a broader distribution so that we clearly correspond to the situation discussed in Sec.~\ref{case2}.
\item $\mu_i =0.1 + {\cal{N}}(0,\sigma_i=1)$: This is the same as 2, but with a smaller deviation from GR as compared to the assumed level of the noise in the detectors. This case breaks the assumptions of Sec.~\ref{case2}.
\end{enumerate}
\begin{figure}[t]
\includegraphics[width=\columnwidth,clip=true]{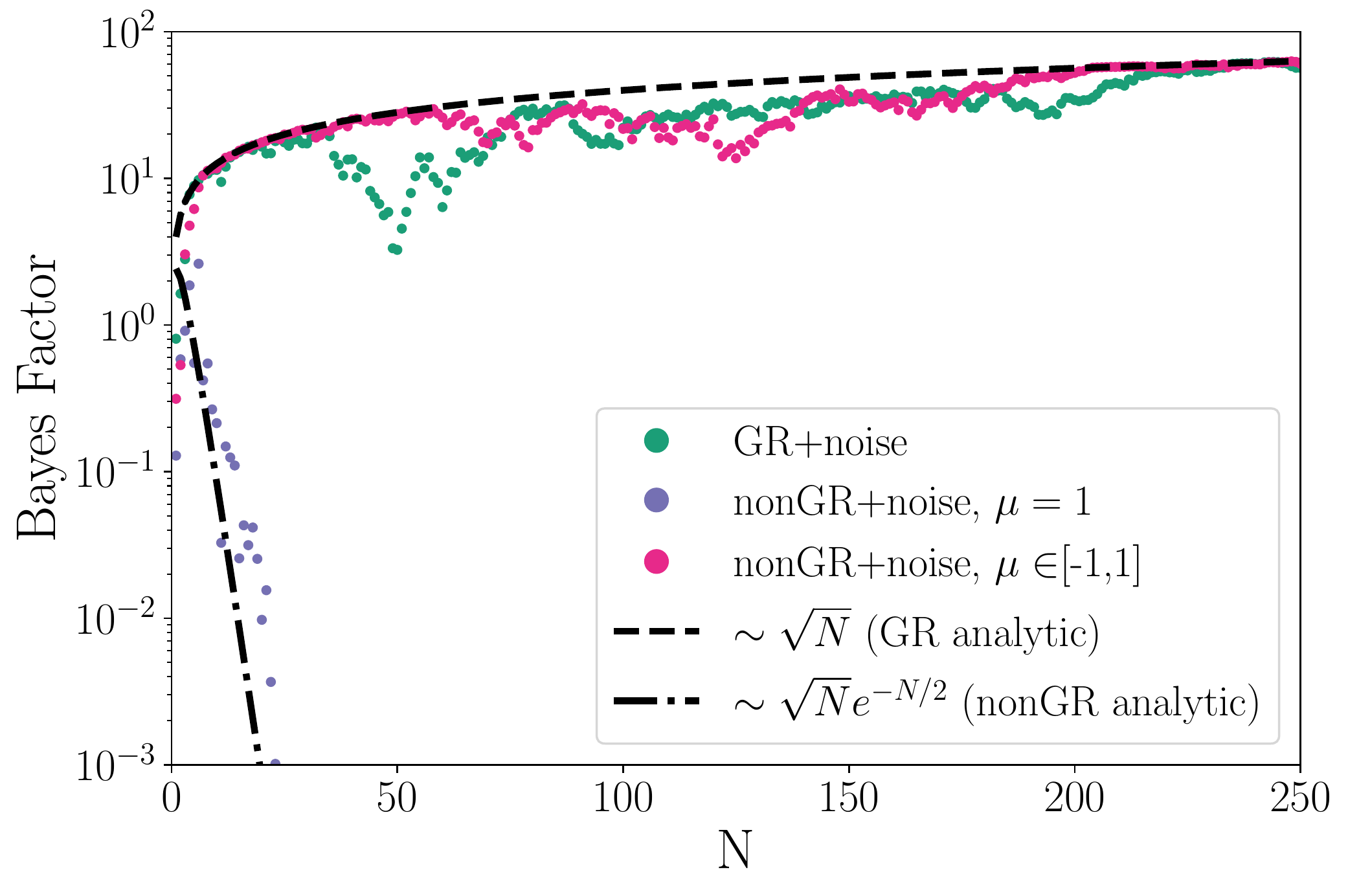}\\
\includegraphics[width=\columnwidth,clip=true]{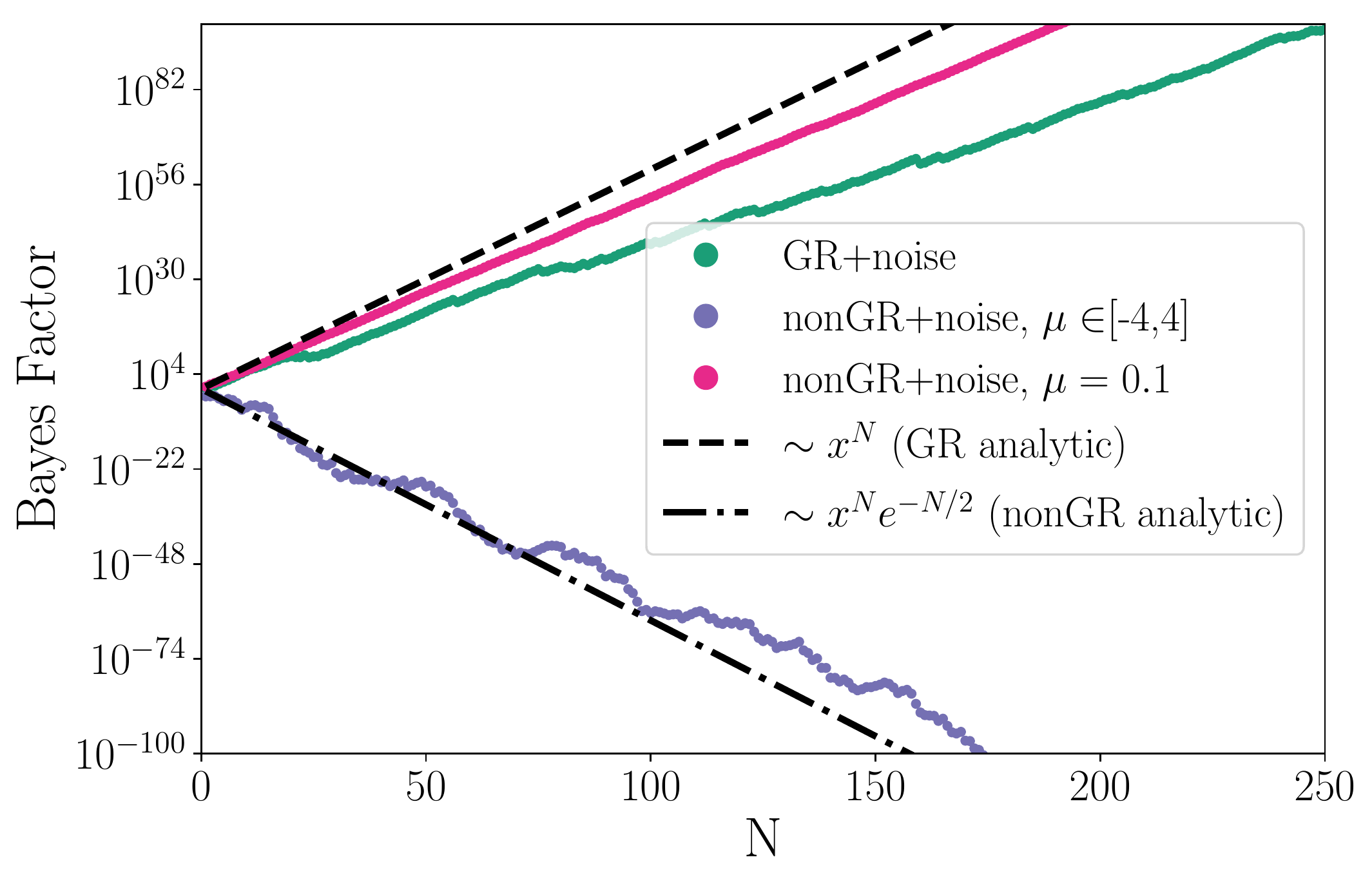}
\caption{ \label{fig:BFs} Combined Bayes Factors as a function of the number of events for an analysis that assumes common $\lambda$ parameters between all events (top panel) and uncorrelated $\lambda$ parameters for each event (bottom panel). 
In each panel the black lines show the theoretically expected scaling of the BF if GR is correct or if it is wrong. 
The purple dots show the BF for a non-GR theory of gravity that obeys the assumption of each BF calculation on $\lambda$, while the pink dots show the BF for a non-GR theory of gravity that does not obey the assumption of each BF calculation on $\lambda$. 
The purple dots show exponentially decreasing BFs, while the pink ones lead to BFs in favor of GR, leading to the incorrect conclusion. 
These results suggest that the two non-GR theories cannot be tested simultaneously in a model-independent way.}
\end{figure}

We then compute the resulting BFs for each scenario of Secs.~\ref{case1} and~\ref{case1} numerically. 
The results are in Fig.~\ref{fig:BFs} whose top panel shows the BF as a function of the number of events computed according to Eq.~\eqref{BFcommon}, while the bottom panel corresponds to the case of Eq.~\eqref{BFrandom}. 
On both panels the black lines show the expected BF scaling for GR and non-GR, in the case of zero-noise realizations.
The green dots refer to the case where GR is the correct theory of gravity but also accounts for the specific noise realization in the detectors. 
In both cases we find that even with noise the combined BFs favor GR as expected, although the exact scaling is affected compared to a set of zero-noise realizations. 

In both panels the purple dots correspond to the BFs computed given a non-GR theory of gravity that obeys the assumptions under which each individual BF expression was derived. 
Specifically, in the top panel we show the second injection case above, in which all GW systems share a common nonzero parameter, while in the bottom panel we show the fourth injection case which corresponds to the scenario where each system has a randomly chosen parameter in the range $[-4,4]$. 
Both these theories lead to sharply decreasing BFs correctly signaling a deviation from GR.

Finally, the pink dots in both plots correspond to non-GR theories of gravity that \emph{break} 
the assumptions under which the BF expression was derived. 
In the top panel, we show the results from the third case, while in the bottom panel we show the fifth case.
Both lead to BFs in favor of GR, and as a consequence we would conclude that GR is correct and fail to detect the deviation. 
Specifically, if the correct theory of gravity predicted parameters distributed symmetrically around $0$ and we analyze these systems assuming they have the same parameter (which is equivalent to multiplying the individual likelihoods), we would erroneously build confidence in favor of the wrong theory of gravity, as shown on the top panel. 
We encounter a similar situation in the right panel, where the pink dots are in fact more in favor of GR than the case where GR is correct and the data is noisy.

The above considerations lead us to a seemingly obvious conclusion: if we perform an analysis under certain assumptions which in reality are violated, we are in peril of obtaining biased answers. 
However in the case of testing GR this statement has additional implications. 
In Sec.~\ref{sec:analytical} we argued that certain assumption \emph{needs} to be made when combining many events, in the form of selecting an expression for the term $p(\lambda_2|\lambda_1)$. 
As a consequence, it is not possible to test all possible theories of gravity with a common analysis, and so there can be no truly model-independent test of GR using multiple GW detections.

\section{Full GW analysis: Measuring the graviton mass}
\label{sec:graviton}

As a concrete example of how a joint analysis, calling on information from several independent observations, can be either powerful or misleading we present the case of a theory where the graviton has an unknown but finite mass.
The emitted GW will then be dispersed as it propagates through the universe.
This will manifest as a modification of the GW phase relative to the GR prediction where the graviton is assumed to be massless.

In the case of a nonzero graviton mass, the waveform will be modified starting at the 1PN order\footnote{In the post-Newtonian framework, an $N$PN order term is a factor of $(u/c)^{2N}$ over the leading order term, where $u$ is some characteristic velocity of the system and $c$ is the speed of light.}~\cite{Will:1997bb} and the GW phase will get an additional leading-order term of
\begin{equation}
\Psi_{g}=-\frac{\pi^2 D {\cal{M}}}{\lambda_G^2 (1+z)}u^{-1},
\label{eq:PsiG}
\end{equation}
where ${\cal{M}}$ is the detector frame chirp mass of the system, $\lambda_G$ is the graviton Compton wavelength (with GR assuming $\lambda_G=\infty$), $z$ is the redshift, and $D$ is
\begin{equation}
D=D_L \frac{1+(2+z)(1+z+\sqrt{1+z})}{5(1+z)^2},
\end{equation}
with $D_L$ being the luminosity distance. 
The frequency dependence of this term is encoded in $u=\pi {\cal{M}}f$.

In the ppE framework~\cite{Yunes:2009ke}, the 1PN ppE parameter $\beta_1$ is nonzero and equal to $\Psi_G$, while all other ppE terms $\beta_i$ vanish. 
Using the redefinition of~\cite{Agathos:2013upa} where the ppE deviations are expressed as relative phase shifts $d\chi_i$, compared to the GR PN term, the only nonzero modification is
\begin{equation}
d\chi_2=\frac{\Psi_G}{\frac{5}{96}\left( \frac{743}{336}+\frac{11}{4}\eta \right)\eta^{-2/5}u^{-1}}.
\label{eq:dhi2}
\end{equation}

Restricting to this specific beyond-GR model allows us to study combining information under different assumptions, as the theory is both well understood and easily quantified through a universal additional parameter $\lambda_G$.

 \subsection{Simulated BBH systems}
  
We simulate $18$ binary black hole (BBH) systems with masses drawn from a distribution between $4-45 M_\odot$, isotropic spin directions, and dimensionless spin magnitudes distributed uniformly between $0-0.8$.
The emitted gravitational waves are described by the IMRPhenomPv2 waveform family~\cite{Hannam:2013oca,Husa:2015iqa,Khan:2015jqa}, with all systems assuming a common graviton Compton wavelength of $\lambda_G^{\mathrm{true}}=10^{12}$ km. 
Notice that while this value of $\lambda_G$ has been ruled out by recent GW observations \cite{Abbott:2017vtc}, here we are primarily interested in an illustration of how a deviation of GR could be extracted from multiple signals, and not to directly show how well the graviton mass can be constrained. 

The BBHs are distributed uniformly across the sky with distances drawn from a distribution uniform in co-moving volume, assuming the cosmology defined in~\cite{Ade:2015xua}.
We inject those systems in a noise-free network of 3 advanced LIGO detectors~\cite{TheLIGOScientific:2014jea} operating at design sensitivity~\cite{aLIGO_oldDesignPSD}, including the currently-operational LIGO-Hanford and LIGO-Livingston detectors, and the under-construction detector LIGO-India.
All of the 18 systems considered here have a network signal-to-noise ratio above 12.
 
We perform parameter estimation on those simulated signals with the publicly-available software library LALInference~\cite{Veitch:2014wba,lalsuite} under the same prior assumptions as the BBH analyses presented in~\cite{LIGOScientific:2018mvr},  with the addition of a ppE-type deviation from GR at the 1PN order parametrized as in Eq.~\eqref{eq:dhi2}, with a prior defined as uniform across $-30 \geq d\chi_2 \geq 30$.
 
 \subsection{General ppE test}

Following Eqs.~\eqref{eq:PsiG} and~\eqref{eq:dhi2}, the specific true value of the parameter $d\chi_2$ differs from source to source as their respective masses and distances vary.
This means that a joint analysis assuming a common value for $d\chi_2$ for all the BBHs will result in incorrect conclusions when applied to the massive graviton model.
We present such an analysis here to illustrate what can occur when ppE parameters are assumed to be common among sets of GW detections.\footnote{Note that published constraints on the graviton mass using multiple GW events \emph{do} correctly combine inferences on the common parameter $\lambda_G$ \cite{Abbott:2017vtc}, contrary to our illustration in this section.}

\begin{figure}[]
\includegraphics[width=\columnwidth,clip=true]{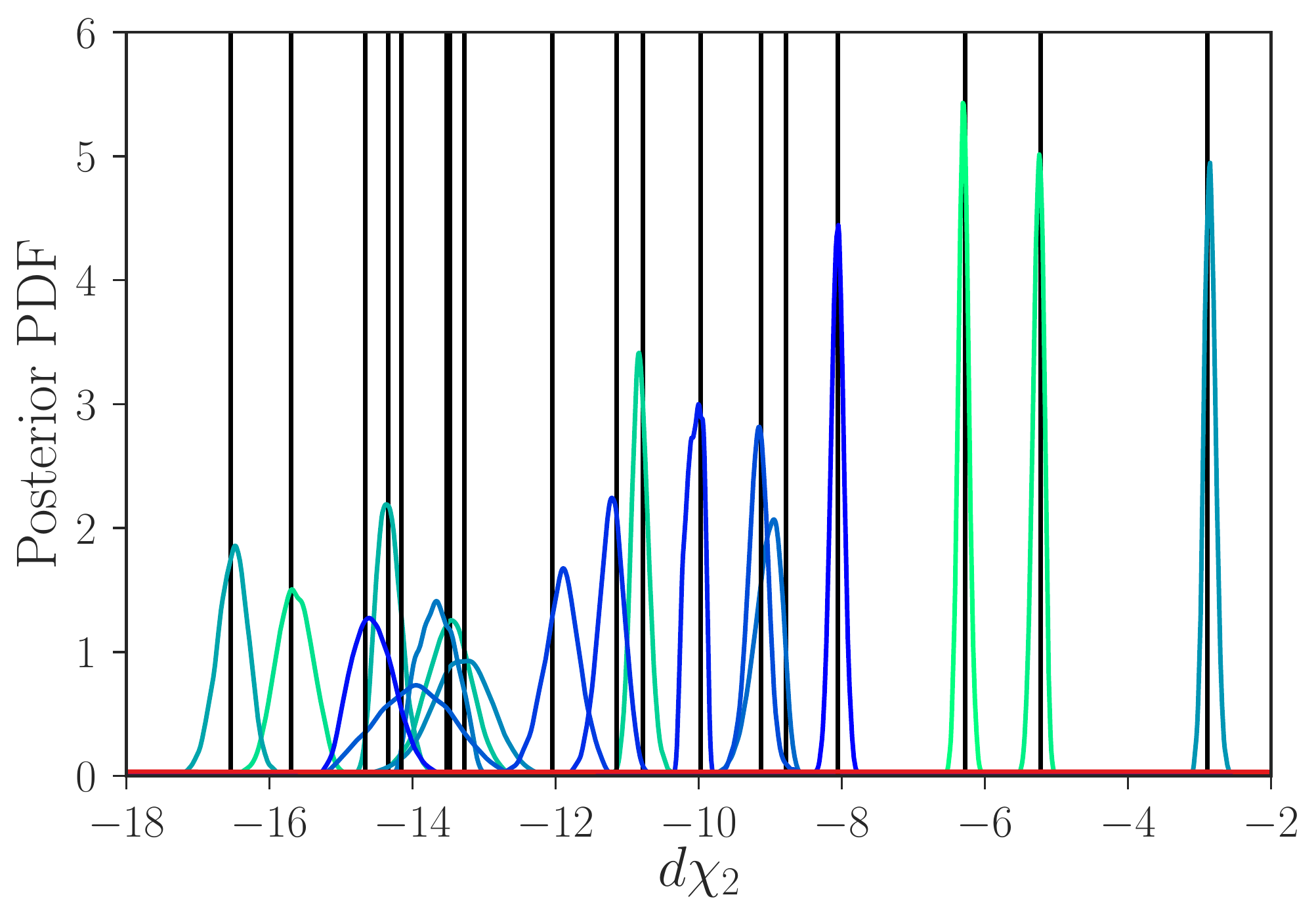}
\caption{\label{fig:dchi2}
Posterior distributions for the $d\chi_2$ parameters measured individually for the 18 BBH events.
The colours of the distributions are only illustrative in this figure, but the transition from green to blue follow the same order as in Fig.~\ref{fig:lambdaG}.
The true $d\chi_2$ for each event is shown as an black vertical line.
The red horizontal line represents the assumed prior probability distribution.
Most $d\chi_2$ distributions are effectively disjoint, and as such any common distribution created from them would lack support throughout the range over which the prior is specified.
}
\end{figure}

In the case where the individual posteriors for $d\chi_2$ from each event are broad and overlap each other, assuming a common $d\chi_2$ and combining likelihoods as in Eq.~\eqref{eq:CommonPosterior} is expected to lead to uninformative joint posteriors.
We would expect to recover the prior on $d\chi_2$, and combining BFs as in Eq.~\eqref{BFcommon} would therefore be uninformative.
Here we face another situation, as illustrated in Fig.~\ref{fig:dchi2}, where we show the posterior distributions for $d\chi_2$ for all our $18$ events. 
Because our chosen $\lambda_G$ provides a relatively strong deviation in the waveforms, each single-event analysis can lead to the conclusion that a deviation from GR is present.
None of the posteriors is consistent with the GR prediction of $d\chi_2 = 0$.
However, many of the $d\chi_2$ posteriors shown in Fig.~\ref{fig:dchi2} are disjoint, so the joint posterior distribution would have no support over its prior range.
In this case it is not possible to make a statement about the nature of the observed deviation, or combine the observations in order to build further evidence of the deviation, without modeling the observed $d\chi_2$ distributions as arising from an underlying population.

Hence, treating the general $d\chi_2$ parameter as universal, and as described in Sec.~\ref{sec:CombSynth}, combining their likelihoods accordingly, will not provide any relevant information about the validity of GR from the ensemble of observations.
 
 \subsection{Graviton mass analysis}

\begin{figure}[]
\includegraphics[width=\columnwidth,clip=true]{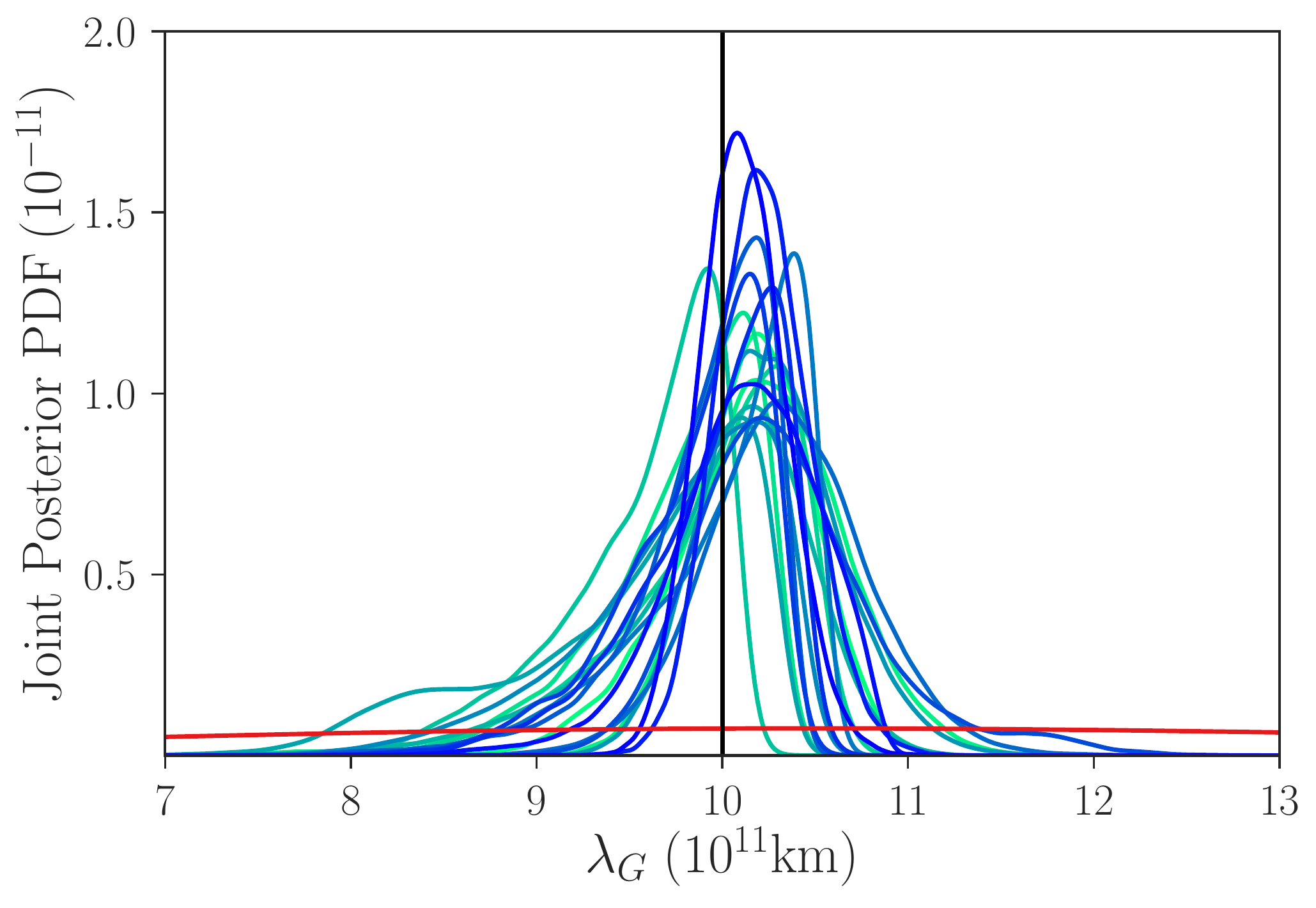}\\
\includegraphics[width=\columnwidth,clip=true]{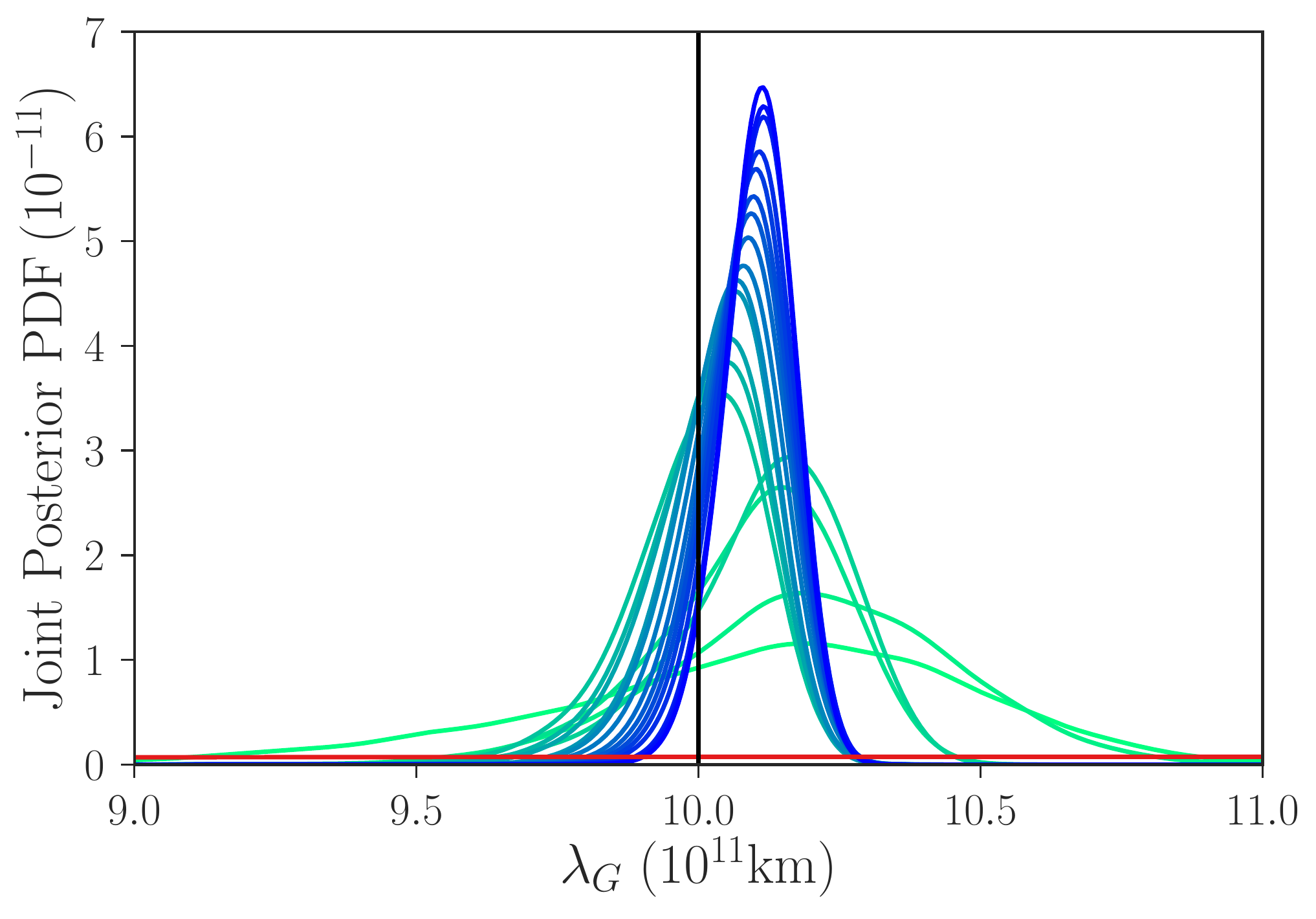}
\caption{\label{fig:lambdaG}
Top: Individual $\lambda_G$ posterior distributions for the 18 BBH events.
Bottom: Joint posterior distributions for the same 18 BBH events.
The colors are assigned such that the green joint distribution contains only one event, and as they transition to blue more events are included.
The distributions shown in the top figure follow the same color transient as the order in which they are added to the bottom joint distribution.
The assumed graviton wavelength $\lambda_G^{\mathrm{true}} = 10^{12}$ km is shown as a black vertical line.
The red near-horizontal line represents the assumed prior probability distribution.
}
\end{figure}

If we instead make the initial, and very strong, assumption that the observed deviations shown in Fig.~\ref{fig:dchi2} are generated by the presence of a massive graviton, we can reinterpret them as describing the fundamental constant associated with this specific extension of GR, in this case the graviton Compton wavelength $\lambda_G$.
For each sample in the posteriors, we compute $\lambda_G$ from $d\chi_2$, $D_L$, our assumed cosmology, and the masses using Eqs.~\eqref{eq:PsiG}--\eqref{eq:dhi2} in order to arrive at a posterior for $\lambda_G$ for each event.
Now the assumptions about creating a joint posterior distribution described in Sec.~\ref{case1} are applicable, since the likelihood distributions from the individual analyses can be multiplied together to constructively build up the available information, as is shown in Fig.~\ref{fig:lambdaG}.

The joint measurement of $\lambda_G$ from these events is $1.0115^{+0.0086}_{-0.0118}\times 10^{12}$ km given as the maximum posterior value and its associated $90\%$ credible interval, which encompasses the true value of $\lambda_G^{\mathrm{true}}=10^{12}$ km.
Fig.~\ref{fig:lambdaG} shows a slight bias towards larger values of $\lambda_G$. This is caused by the fact that the distance posteriors are relatively uninformative for these signals, and largely follow the assumed prior distribution where $p(D_L) \propto D_L^2$. Our inferences on $\lambda_G$ are then affected due to the fact that $\lambda_G \propto \sqrt{D_L/d\chi_2}$.
For a larger set of observations, with a wider spread in both their source parameters and their SNR distribution, this bias is expected to diminish.
 
 \section{Conclusions}
 \label{sec:conclusions}

In this work, we revisit two common methods for combining information from multiple sources employed in the testing-GR literature
and argue that they are the limiting cases of a more generic hierarchical framework. 
Moreover, we show that both methods make certain assumptions about the underlying theory of gravity being tested.
In particular, multiplying the likelihood functions of beyond-GR parameters from individual binary merger events 
amounts to imposing that all signals share the same beyond-GR parameter.
Meanwhile, computing the BF in favor of GR from each event and multiplying the BFs together makes the implicit 
assumption that each binary has an additional beyond-GR parameter, and these parameters are unrelated to
each other.

We argue that both assumptions are expected to fail for generic theories of gravity. We present numerical results of BFs
where we show that if the true theory of gravity does not obey the assumptions under which information was
combined and the BF was calculated, we might fail to detect a deviation from GR.

Our results suggest that combining inferences from multiple sources cannot be performed in a completely
model-independent way such that it applies to generic theories of gravity. Even in the generic
hierarchical analysis, a model for the population distribution needs to be specified in some fashion.
We present here the simple case where the beyond-GR parameters are distributed according to a normal 
distribution, though with enough detections multiple models for the common distribution can be explored and 
compared, while their hyperparameters are measured.

We have also applied these ideas to a more realistic study of signals generated assuming a theory where the graviton is massive, which we have analyzed using a full parameter estimation analysis.
We recover the signal parameters in two ways: either we make inferences on the Compton wavelength of the graviton directly, in which case the deviation parameter is the same among all the signals, or we use a generic ppE parameterization where the beyond-GR parameter differs from signal to signal. 
Although for illustration we select a graviton mass where the deviations from GR are detectable in the individual signals, we argue that when using incorrect assumptions on how to combine these signals we would fail to detect a deviation from GR in the joint analysis. 
Meanwhile, using the correct method for combining the posteriors allows us to draw tight constraints on the graviton mass using the full dataset.
This exercise highlights the needs to re-interpret the ppE parameters in terms of specific theories of gravity 
before combining measurements of them.

We emphasize that even though we frame our results in the context of testing for the true theory of gravity,
 our conclusions are applicable to any problem that involves combining information, for example in making cosmological measurements and in inferring populations of compact binaries, where these ideas have long been in use.
 
 \acknowledgements

We thank Will Farr, Max Isi, Walter Del Pozzo, and Salvatore Vitale for helpful discussions.
We are grateful for computational resources provided by Cardiff University, and
funded by an STFC grant supporting UK Involvement in the Operation of Advanced
LIGO. 
C.-J. H. acknowledge support of the MIT physics department through the Solomon Buchsbaum Research Fund, the National Science Foundation, and the LIGO Laboratory.
The Flatiron Institute is supported by the Simons Foundation.
This is LIGO Document Number P1900098.

\bibliography{OurRefs}

\end{document}